\documentstyle[epsf]{l-aa}

\newcommand{\aaa}{A\&A}      
\newcommand{\apj}{ApJ}       
\newcommand{\mnras}{MNRAS}   
\newcommand{\bm}[1]{\mbox{{\boldmath $#1$}}}

\newcommand{\fr}[2]{\frac{\displaystyle #1}{\displaystyle #2}}

\newcommand{\pder}[3]{\fr{{\partial}^{#3} {#1}}{{\partial} {#2}^{#3}}}

\newcommand{\cross}{\times}
\renewcommand{\b}{\!\!\!}
\newcommand{\erg}{{\rm \ erg}}

\newcommand{\cm}{{\rm \ cm}}
\newcommand{\kpc}{{\rm \ kpc}}
\newcommand{\pc}{{\rm \ pc}}   %
\newcommand{\s}{{\rm s}}
\newcommand{\K}{{\rm \ K}}

\newcommand{\Myr}{{\rm \ Myr}}
\newcommand{\G}{{\rm G}}

\begin{document}

\thesaurus{02(02.13.1; 02.09.1; 11.13.2; 11.19.2; 11.09.4; 09.11.1)}

\title{The galactic dynamo effect due to Parker-shearing instability of
magnetic flux tubes}
\subtitle{III. The fast dynamo model.}

\author{M. Hanasz \inst{1} \and H. Lesch \inst{2}}

\institute{Centre for Astronomy, Nicolaus Copernicus University,
PL-87-148 Piwnice/Toru{\'n}, Poland, ({\em mhanasz@astri.uni.torun.pl})
\and
University Observatory, M\"unchen University,
Scheinerstr. 1, 81679 M\"unchen , Germany}

\offprints{M. Hanasz }

\date{Received 4 July 1997/ accepted 12 December 1997}

\maketitle

\markboth{M.  Hanasz et al.: The galactic dynamo effect due to Parker-shearing
instability of magnetic flux tubes }{}

\begin{abstract}
We present a new fast dynamo model for galactic magnetic fields, which is based
on the Parker-shearing instability and magnetic reconnection, in the spirit of
the model proposed by Parker (1992).  We introduce a new scenario of flux tube
interactions and estimate the dynamo transport coefficient basing on simple
geometrical arguments.  The obtained expressions are equivalent to the formally
derived helicity $\alpha_d$ and diffusivity $\eta_d$ in the first paper of this
series.  The model we propose predicts that the $\alpha$-effect in galactic
discs has opposite sign with respect to that resulting directly from the sign
of the Coriolis force.  We estimate the rate of magnetic heating due to the
reconnection of magnetic flux tubes, which plays an important role in our
dynamo model.  The corresponding luminosities of the diffuse X-ray emission are
consistent with the ROSAT observations of nearby galaxies.

The present considerations synthesize  the ideas of Parker with our own results
presented in the preceding papers (Hanasz \& Lesch 1993, 1997; Hanasz 1997).


\end{abstract}

\section{Introduction}

The 'classical' galactic dynamo mean field theory based on Kolmogorov
turbulence experienced a fundamental criticism by Kulsrud \& Anderson (1992)
and Vainshtein \& Cattaneo (1992).  It was argued that the classical mean-field
dynamo approach misses the magnetic back reaction on the turbulent flow.
Thereby the turbulent diffusion coefficient is reduced as well as the turbulent
helicity.  The effect of Kolmogorov turbulence on the evolution of a weak
large-scale field was found to build up small scale magnetic fields on the
equipartition level very quickly and then begins to dissipate by ambipolar
diffusion (Kulsrud and Anderson 1992).

Parker (1992) (hereafter P' 92) reviews the application of dynamo theory to
galactic magnetic fields, he argues like Vainshtein and Cattaneo (1992) that a
weak large scale magnetic field grows to a certain limit, which is very
small compared to the equipartition magnetic field.  Above this limit the
cascade of magnetic energy toward diffusive scales is blocked by the Lorentz
force and turbulent diffusivity as well as the $\alpha$-effect are reduced.
However, unlike Kulsrud and Anderson (1992) and Vainshtein and Cattaneo (1992),
he does not draw the conclusion that the galactic magnetic field must be of
primordial origin.  Instead, he argues that the agreement of the predictions of
dynamo theory with observations suggests that it is basically sound, so that
some other explanation should be found for the relative large values of the
turbulent diffusivity that seem to give consistent results.

Parker's model of buoyancy driven galactic dynamo is as follows:  Since the
interstellar medium is composed of the ionized gas, magnetic field and cosmic
rays, the last two weightless components support the heavy gas against vertical
gravitation due to the stellar disc.  Such a configuration is intrinsically
unstable against vertical perturbations of initially azimuthal magnetic field
lines.  The ionized gas slips down along magnetic field lines forming gas
condensations in valleys and the cosmic ray gas tends to escape from the
galactic disc together with an amount of the trapping magnetic field.  The main
effect of cosmic ray gas is to inflate the raised region and form magnetic
lobes (see Fig.~2 of P'92).  The $\Omega$-shaped loops form and are tightly
packed due to the inflation.  At the boundary of adjacent lobes the magnetic
field lines of opposite direction are pressed together and a fast magnetic
reconnection starts to rearrange the magnetic field configuration.  As a
result, close magnetic loops form which are free to rotate since magnetic
tension does not counteract rotation.  Then, further reconnection fuses many
loops to form a large scale poloidal magnetic field and galactic differential
rotation forms the azimuthal field.  The reconnection at current sheets at the
boundary of lobes is supposed to transform an amount of magnetic energy into
heat that maintains the $10^6-10^7$ K temperature of the halo gas producing the
halo X-ray emission.  Since the magnetic reconnection is one of the
simultaneously working heating processes, the observed X-ray luminosity of the
order of $10^{40}$ erg/s determines an upper limit of magnetic reconnection
heat output.  Parker postulates that such a process can be described by the
conventional $\alpha\omega$-dynamo equations.

The term ``fast dynamo'' refers to the dynamos for which the generation rate
of the mean magnetic field remains finite in the limit of vanishing
resistivity.  The resistivity is responsible for the dissipation of the
small scale magnetic field.  In the galactic conditions the Kolmogorov cascade
is cut off by viscosity at scales much larger than the resistive scale.  Then,
in the standard dynamo theory the resistive dissipation is suppressed, which
implies that the  transition from chaotic to the uniform components of
magnetic field becomes infinitely slow.  In the Parker's fast galactic dynamo
the transition between small and large scale magnetic fields is possible
(within a finite time) in the limit of vanishing (i.e.  enormously small)
resistivity because of the presence of tangential discontinuities in magnetic
field.

The magnetic reconnection which takes part in the dissipation process of fast
dynamos is supposed to proceed as fast as the Alfven speed $v_A$ divided by
logarithm of the Lundquist number $N_L = l v_A/\eta \sim 10^{15} - 10^{20}$.

Our aim is to modify some elements of the Parker scenario and provide some more
details according to our detailed studies of the dynamo effect due to the
Parker-shearing instability contained in Papers I and II.  The key points of
the Parker's scenario are based on the following facts:

\begin{enumerate}
\item The ascending magnetic flux tubes inflate vigorously and the closely
packed magnetic lobes form and reconnect to form closed loops of magnetic
field.
\item There are two patterns of reconnection illustrated in Fig.~3 of P'92:
(a) at the bases of lobes and (b) at the upper regions of the lobes. The second
pattern is suggested to be more likely.
\item In both the patterns reconnection
operates on single magnetic flux tubes, otherwise it would not form closed
loops.
\item The closed loops are free to rotate because they are disconnected from
the surrounding and therefore released from the restoring magnetic tension.
\end{enumerate}

We would like to propose a slightly modified picture following from our
calculations of Papers I and II:
\begin{enumerate}
\item The magnetic tension limiting cyclonic deformations (i.e. these which
provide the helicity $\alpha_d$) is compensated by two effects: (i) the
powerful contribution of cosmic rays in buoyancy, which is free of restoring
tension and (ii) the differential forces counteracting the magnetic tension,
due to the rotational shear and the density waves.  As a result, even open
magnetic field lines produce a strong $\alpha$-effect in the nonlinear regime
of the Parker-shearing instability.

\item The related radial displacement amplitudes of magnetic flux tubes are in
general comparable to vertical ones and exceed the cosmic ray driven
thickening of the lobes. This implies that adjacent magnetic flux tubes are
predominantly pressed together by the horizontally twisting Coriolis force.
\item Due to this fact the reconnection should take place in two locations on
the flux tube which are maximally displaced in the radial direction due to the
Coriolis force.
\item The most likely possibility is such that reconnection
operates on  some
neighboring flux tubes and does not form closed loops from a single flux tube.
\end{enumerate}

The above remarks allow us to propose a geometrically different scenario, which
remains still in the spirit of the Parker's model. In the present
approach the classical concept of turbulence is replaced by more or less
chaotic superposition of unstable Parker modes.  The perturbations may be
uncorrelated along the mean direction of magnetic field over distances of the
order of a few wavelength of the Parker instability.  We assume additionally
that neighboring flux tubes can be braided and move individually or coherently
within bundles.

We shall estimate the heat output from magnetic reconnection in our model and
show that this is in a general agreement with the X-ray data which suggest that
an amount of diffuse X-ray emissivity of non stellar origin in some nearby
spiral galaxies.  We shall take NGC 6946, M51 and NGC 1566 as examples.

\section{The new dynamo model}

\subsection{Geometry of flux tube interactions}

The pattern of deformation of flux tubes due to the Parker-shearing instability
projected on the planes $(\phi,r)$ and $(\phi,z)$ is shown in Fig.~1.  We show
only single periods of perturbation of two distinct azimuthal flux tubes (plus
small pieces of two other flux tubes), characterized by the same magnetic flux
$\Phi$.  Let us suppose for simplicity that the initial height above the
galactic midplane is $z_0$ for both the flux tubes, and the vertical
displacement amplitude due to the Parker-shearing instability is $\Delta z$.
The initial radial separation of the flux tubes is $2\Delta r$ and the radial
displacement amplitude is $\Delta r$.

\begin{figure}
\epsfxsize=\hsize \epsfbox{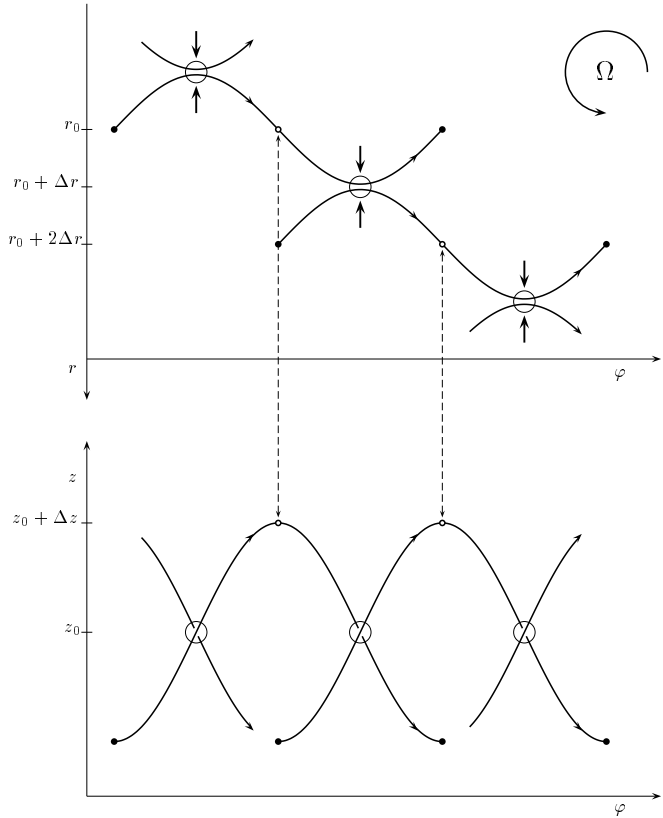}
\caption[]{
The horizontal and vertical deformations of two distinct flux tubes.
The places of reconnection correspond to the radially most
displaced points of the flux tubes and are enclosed by circles. }
\end{figure}

\begin{figure}
\epsfxsize=\hsize \epsfbox{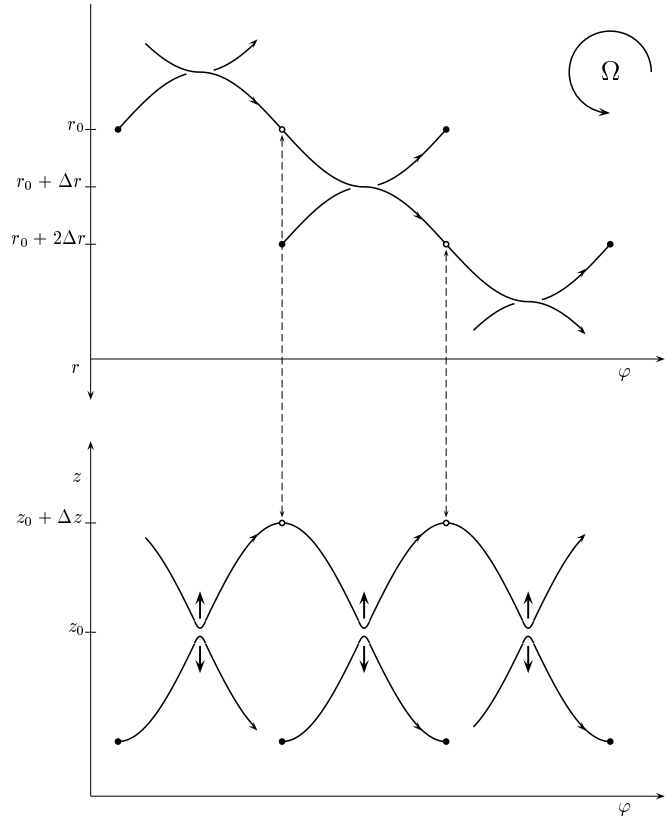}
\caption[]{The final configuration of magnetic flux tubes after reconnection.
The flux tubes belong to two families, one remains in the disc and the second
is shifted toward the galactic halo. Both of them contain the radial component
of magnetic field.}
\end{figure}

The scenario, we propose is as follows.  Due to the action of the
Parker-shearing instability the flux tubes undergo cyclonic deformations as
shown in Fig.~1.  The horizontal deformations press different flux tubes
together in places enclosed by circles, which are maximally displaced in the
radial direction.  The flux tubes form current sheets at their boundary due to
the magnetic field gradient and reconnection starts to rearrange and to relax the
configuration of magnetic field.
The new configuration of magnetic flux tubes, is shown in Fig.~2.

Two new families of flux tubes appear replacing the original azimuthal magnetic
field.  The first family is formed from the lower parts of the original flux
tubes and remains in the galactic disc.  The second family formed from the
upper parts shifts toward to the galactic halo, since the tops of the flux
tubes are typically displaced by the vertical scale height $H$  of ionized
gas.  Both families acquire a specific amount of radial magnetic field, which
can be estimated basing on simple arguments, without invoking the complex
techniques of computing the $\alpha_d$ coefficient.

Let us apply the linear approximation (see Paper I) to describe the
perturbations on the initially azimuthal magnetic flux tubes in the
rectangular reference frame $(x,y,z)$ locally representing the azimuthal,
radial and vertical coordinates, respectively. The radial and vertical
displacements are
\begin{eqnarray}
y = Y \exp (i k s) \exp (\omega_i t),\\
z = Z \exp (i k s) \exp (\omega_i t),
\end{eqnarray}
where $s$ determines the position along the flux tube, $Y$ and $Z$ are the initial
displacement amplitudes, $k$ and $\omega_i$ are the wavenumber and the growth
rate, respectively. The corresponding Lagrangian velocities are
\begin{eqnarray}
v_y = \pder{y}{t}{} =\omega_i z \frac{Y}{Z}, \ \ \
v_z = \pder{z}{t}{} = \omega_i z,\\
\end{eqnarray}

Assuming that perturbations on two flux tubes displaced by $\Delta r$ start
to grow simultaneously we can estimate the time interval necessary for a collision
of the two flux tubes
\begin{equation}
\Delta t \simeq \frac{\Delta r}{v_y}
= \frac{\Delta r}{\omega_i Y/Z \Delta z}
\end{equation}
If we denote the wavelength of the perturbation by $\lambda$, then after
reconnection each family of magnetic flux tubes has a component of radial
magnetic field
\begin{equation}
\Delta B_r \simeq \frac{4 \Delta r}{\lambda} B_{\varphi}
= \frac{2}{\pi} k \Delta r B_\varphi
\end{equation}
where $B_\varphi$ is the initial azimuthal magnetic field. The radial magnetic
field is generated at the rate
\begin{equation}
\frac{\Delta B_r}{\Delta t} \simeq \frac{2}{\pi} \omega_i k \Delta z
\frac{Y}{Z} B_\varphi
\end{equation}
The coefficient on the rhs. in front of $B_\varphi$ is quite similar to the
analytical expression for $\alpha_d$ (Eqn. (85) in Paper I). Since in the thin
disc approximation $\alpha_d$ relates $\partial B_r/\partial t$ with
$\partial B_\varphi/\partial z$ we should take into account the assumed
stratification of the interstellar medium which according to formulae (18)-(21)
of Paper I gives
 \begin{equation}
B_\varphi = -2 H \pder{B_\varphi}{z}{} \label{Bstrat}
\end{equation}
We obtain finally
\begin{equation}
\frac{\Delta B_r}{\Delta t} \simeq \frac{4}{\pi} \omega_i k \Delta z H
\frac{Y}{Z} \pder{B_\varphi}{z}{}
\end{equation}
For perturbations with the vertical amplitude $\Delta z \sim H$ we obtain
the coefficient on the rhs
\begin{equation}
\alpha_d = \frac{4}{\pi}\omega_i k \frac{Y}{Z} H^2
\end{equation}
which is consistent with the formula (85) of Paper I with the accuracy to the
numerical coefficient of the order of 1.

The vertical amplitudes of perturbations of the order of the vertical
scale height $H$ imply that the upper family of magnetic flux tubes forms at
the base of the galactic halo and is in fact lost for the disc.  This can be
interpreted as a kind of diffusion of the magnetic field from the disc.
The escape rate of magnetic field from the disc to halo can be estimated
from the Poynting flux in vertical direction

\begin{eqnarray}
S_z &=& \frac{c}{4 \pi}({\bm{E} \cross \bm{B}})_z\\
    &=& \frac{1}{4\pi}(v_z B_{\varphi} - v_{\varphi}
        B_z)B_{\varphi}.\nonumber
\end{eqnarray}
Without going into details we can estimate the Poynting flux through a
horizontal boundary between disc and halo

\begin{equation}
S_z \simeq \frac{1}{2}\frac{1}{4\pi} v_z B_{\varphi}^2 \label{S_z}.
\end{equation}
The additional numerical factor $1/2$ in (\ref{S_z}) is because only half
of wavelength of the flux tube is ascending and transports the magnetic energy
through the horizontal plane at the boundary of disc and halo. We express
the conservation of magnetic energy
as if the other components, i.e. thermal gas and cosmic rays, were absent
\begin{equation}
\pder{}{t}{} \left(\frac{B^2}{8\pi} V \right)   = 2 S_z A_{disc}, \label{emag}
\end{equation}
where $V = 2 A_{disc}\cdot H$ is a volume of a disc with surface area
$A_{disc}$ and half
widths H. The numerical factor $2$ in
(\ref{emag}) is because we consider two surfaces at $z=\pm H$ enclosing the
disc.   With the assumed approximate equipartition of thermal, magnetic
and cosmic ray energies, the error we make is at most of the order of 1.  The
above energy conservation equation leads to the following equation describing
loses of magnetic field due to the vertical motions driven by the
Parker-shearing instability

\begin{equation}
\pder{\bm{B}}{t}{} = \frac{v_z}{2 H} \bm{B}.
\end{equation}
Taking the relation  (\ref{Bstrat}) into account we obtain
\begin{equation}
\pder{\bm{B}}{t}{} = 2 v_z H \pder{\bm{B}}{z}{2}, \label{diffterm}
\end{equation}
what corresponds to the turbulent diffusion term in the dynamo equation.
With the accuracy to a numerical factor of the order of 1 the coefficient
on the rhs. of (\ref{diffterm}) is
\begin{equation}
2 v_z H \sim \omega_i Z(t)^2 = 4 \eta_d
\end{equation}
if the vertical displacements due to Parker-shearing instability are of the
order of the vertical scaleheight $H$. Thus, basing on the Poynting flux we
obtained an effect equivalent to the turbulent diffusion in the dynamo
equation.

The above results mean that the presented processes can be described by the
conventional dynamo equation, but the physical interpretation of the dynamo
transport coefficients is now different from the conventional one.

Following the scenario presented in Figs.~1 and 2 we note that the sign
of $\alpha$-effect is positive for the upper family (halo) and negative for the
lower family (disc).  This is because the magnetic field lines turned by
Coriolis force are advected to the halo and lost from the disc.  What remains
in disc after reconnection is like rotated opposite to the Coriolis force.
This would be a possible explanation of the negative $\alpha$-effect measured
by Brandenburg et al.  (1995) in their numerical simulations.

The above properties of our model may be also related to some
observational features of galactic magnetic fields. The magnetic field
structure is typically a spiral, which is parallel with good accuracy to the
optical spiral structure.  In typical galaxies like M51 and
NGC 6946, representing trailing spirals, gas flows into arms from the direction
of inner edges of arms.  The rotation of gas is faster than the rotation of
spiral pattern inside the corotation radius.  For gas streaming from the top of
a rising magnetic loop the cyclonic rotation due to the Coriolis force is
opposite to the galactic rotation, which implies that the upper family of
magnetic field lines tends to acquire a pitch angle of opposite sign with
respect to the pitch angle of spiral arms.  Compare eg.  Fig.~1 of this paper
with the picture of M51 in Fig.~1 of Berkhuijsen et al.  (1997).  The signs of
galactic rotations are the same in both cases, but the upper (halo) family of
magnetic field lines in our case is transversal with respect to the spirals in
the picture of of M51.  The problem is more complicated however, because the
main effect which determines the pitch angle of magnetic field lines is the
differential rotation (see Beck et al.  1996 and references therein).  Without
taking into account of the $\alpha$-effect the differential rotation of the
trailing spiral provides the same sign of the magnetic pitch angle as the pitch
angle of spiral arms.  Thus, in our model the $\alpha$-effect and the
differential rotation operate accordingly on the lower (disc) family of
magnetic field lines.

The situation in halo is different. The signs of the pitch angle resulting
from the differential rotation and the $\alpha$-effect tend to be opposite.
We note that even if on average the effect of differential rotation is
dominating, we can probably expect in galactic halos departures from this
rule.  First of all, it is reasonable to expect that the azimuthal shear
is weaker in galactic halos then in discs.  Second, in contrast to the
differential rotation, the alpha effect may be locally significantly different
than its mean value. This is because the $\alpha$-effect depends essentially on
the cosmic ray pressure (see Papers I and II), which is probably highly
variable depending on position in the disc and the phase of density wave.
A local excess of cosmic rays can easily enhance the $\alpha$-effect by a
factor of 10 or more, together with the substantial increment of the growth
rate of the Parker-shearing instability.

It is possible that in some cases the strong $\alpha$-effect (positive
in halo) is related to the rotation of magnetic field lines toward the
direction transversal to spiral arms.  Then the differential rotation can make
the upper (halo) component of magnetic field even more transversal or possibly
antiparallel to the magnetic field in the disc.  This seems to be observed by
Berkhuijsen et al.  (1997). It is also possible that the described
effects can be responsible for rapid variations and reversals of the
magnetic pitch angle.  One should remember, however, the halo magnetic field is
typically much weaker and has a smaller influence on the radio continuum image
of magnetic structure.  What is usually observed is the stronger disc magnetic
field.

The negative $\alpha$-effect in the disc corresponds to positive dynamo
numbers $D=R_\alpha R_\omega$.  The positive dynamo numbers lead to oscillatory
solution of the dynamo equation in the thin disc approximation.  These
solutions propagate as spiral waves in the disc plane and can couple to the
density waves (Chiba and Tosa 1990; Hanasz, Lesch and Krause 1991, Hanasz and
Chiba 1994, Mestel and Subramanian 1991, Subramanian and Mestel 1993, Moss
1996) if the propagation velocity of the magnetic wave is the same as the
propagation velocity of the spiral structure.  The coupling by means of the
parametric resonance provides an additional growth rate to the dynamo waves.
Since, in the standard dynamo models the dynamo numbers of the real galaxies
are negative, it has been problematic how the corresponding solutions can
couple to density waves in the thin disc approximation.  It has been known,
however, that in the thick discs the oscillatory solutions exist even for
negative dynamo numbers (Starchenko and Shukurov 1989).  The present dynamo
model provides a new justification for the oscillatory dynamo modes and their
coupling to density waves.

In order to close the dynamo cycle we describe next the dissipation process
of the small scale components of the field by magnetic reconnection,
estimate the heat output and compare it with the observed X-ray luminosity
which is a measure for plasma contents in galactic halos.

\subsection{Magnetic heating}

Most of the interstellar plasma is highly conductive, i.e.  the electrical
conductivity is very large (details see below) and any relative motion between
the conductor and the field induce electric currents.  The relation between the
field and plasma is given by Ohm's law ${\bf j}=\sigma \left( \frac{1}{c}{\bf v
\times B} \right) $ which gives for a highly conducting medium (neglecting
external electric fields E, with the current density j and the electrical
conductivity $\sigma$) an infinitely large current density for an infinite
conductivity.  However, this frozen-in property of the magnetic field changes
drastically if field lines with different directions approach each other.  The
chain of processes which is triggered by antiparallel field lines is called
{\bf magnetic reconnection} (e.g.  Biskamp 1994).

Magnetic reconnection is a fundamental intrinsic property of agitated
magnetized, turbulent plasmas (e.g.  Schindler et al.  1991).  Whenever,
magnetic fields with different field directions encounter each other, the
magnetic energy is rapidly dissipated, by accelerating particles and by plasma
heating.  As mentioned above, outside the forming current sheet the motion of
the plasma is ideal, i.e.  the magnetic field lines follow the plasma motion as
if they were frozen into it, which is provided by the high electrical
conductivity.  When the oppositely directed field lines following the plasma
motion approach, the field gradient steepens and the current density
${c\over{4\pi}}{\bf \nabla\times B}$ increases until strong dissipation sets
in.  The problem of reconnection is to know how the dissipation of the currents
with the density $j=e n_{\rm e} v_{\rm d}$ is provided ($n_{\rm e}$ is the
electron number density per ${\rm cm^3}$ and $v_{\rm d}$ denotes the drift
velocity of the electrons relative to the protons).  Rapid dissipation follows
because of the small thickness of the transition layer from one field to
another, providing a thin intense current sheet.  The dissipation may be
further increased when the current density exceeds a critical current density,
i.e when the drift velocity exceeds some thermal velocity of the plasma
constituents.  If $v_d \ge v_{crit}=v_{thermal}$ plasma instabilities excite
plasma waves.  The wave-particle interaction replaces the Coulomb interaction,
thereby increasing the electron scattering frequency (e.g.  Zimmer et al.
1997).

Magnetic reconnection is unavoidable in turbulent plasmas, therefore the
encounter of two antiparallel field components in the ``magnetic atmosphere''
of a galaxy transfers the plasma kinetic energy into heat via compressed,
strained, torn or even decomposed magnetic field structures, which heavily
dissipate the stored magnetic energy in current sheets via magnetic
reconnection.

Magnetic reconnection corresponds to the dissipation of electric currents and
the dissipation (or heating) rate $Q$ (in $\erg \cm^{-3} \s^{-1}$) is

\begin{equation}
Q={j^2\over \sigma}
\end{equation}
With the advent of plasma turbulence
\begin{equation}
\sigma = 2\cdot 10^7 \left[\frac{T}{1 \K}\right]^{3/2} \s^{-1}
\end{equation}
Dissipation is equivalent to either increased current density and/or
reduced electrical conductivity $\sigma$

\begin{equation}
\sigma={\omega_{\rm pe}^2\over{4\pi \nu_{\rm coll}}}.
\end{equation}
where
\begin{equation}
\nu_{\rm coll}\approx \nu_{\rm LH}\approx 4\cdot 10^5
\left[\frac{B}{1 \rm G}\right] \s^{-1}
\end{equation}
is the collision frequency driven by lower hybrid waves with frequency
$\nu_{\rm LH}$, which has
the lowest threshold velocity for the critical current of the order of the
thermal velocity of the ions (e.g. Papadopoulos 1979).
\begin{equation}
\omega_{\rm pe}\sim 5.6\cdot 10^4 \sqrt{\frac{n_{\rm e}}{1 \cm^{-3}}} \s^{-1}
\end{equation}
is the electron plasma frequency. The conductivity and the anomaleous
resistive diffusion coefficient are respectively
\begin{equation}
\sigma\approx 0.6\cdot10^6 \left[ \frac{n_{\rm e}}{0.01 \cm^{-3}} \right]
   \left[\frac{B}{10 \mu \G}\right]^{-1} \s^{-1},
\end{equation}
\begin{equation}
\eta = \frac{c^2}{4\pi\sigma} \approx 1.3\cdot 10^{14}
\left[ \frac{n_{\rm e}}{0.01 \cm^{-3}} \right]^{-1}
   \left[\frac{B}{10 \mu \G}\right] \frac{\cm^2}{\s}.
\end{equation}
The Lundquist number (the magnetic Reynolds number with respect to $v_A$) is
\begin{equation}
N_L = \frac{L v_A}{\eta} \approx 5\cdot 10^{12} \left[\frac{L}{10 \pc}\right]
\left[ \frac{n_{\rm e}}{0.01 \cm^{-3}} \right]^{1/2}
\end{equation}
where $L$ is the assumed typical diameter of colliding flux tubes.

The obtained value of $N_L \sim 5\cdot10^{12}$ suggests that the Parker-Sweet
reconnection pattern with the inflow speeds of the order of
$v_{in} \sim v_A/\sqrt{N_L}$ would be very inefficient and not applicable to
galactic conditions. Even incorporation of the anomaleous resistivity
does not help very much to depart from the limit of large Lundquist numbers.
The Petschek's model, on the other hand appeared to be not valid in the
astophysically relevant range of small $\eta$. It can be used however, as a
phenomenological approach assuming an anomaleous resistivity enhancement in the
diffusion region, such that $R_M \equiv v_{in} L/\eta = M N_L {\cal O}(1)$, but
again, even with the anomaleous resistivity the last condition is not
fulfilled.

Fortunately, the numerical simulations (see eg.  Biskamp 1994) of the
reconnection process lead to the reconnection rates sufficiently large and
independent of resistivity in the small resistivity limit.  This is
accomplished by a nonstationary behaviour in the reconnection region due to the
tearing instability which subsequently leads to small-scale turbulence and
formation of many small-scale secondary current sheets with short life times.
Finally it seems that the resulting reconnection rate is
not much different than $\sim v_A/ \ln N_L$ applied by Parker in his
``fast dynamo'' model.

We give a qualitative estimate of magnetic energy conversion into heat via
magnetic reconnection and rough estimates of the heating process.  The details
of the plasma processes involved have been generally discussed by Lesch (1991)
and for the case of the halo of the Milky Way by Zimmer et al.  (1997).
We ask what magnetic field strength $B_{\rm eq}$ is necessary to heat up a
plasma to X-ray temperatures of several $10^6$K and compare the values with the
observed ones in the halo of the Milky Way (e.g.  Kazes et al.  1991) of about
$10\, \mu$G.  In Fig.~3 the magnetic field

\begin{equation}
B_{\rm eq}=\sqrt{8\pi n_{\rm e} k_{\rm B} T_{\rm e}}
\end{equation}
is computed for a temperature of $2\cdot 10^6$K and
a density range from $0.1\, \cm^{-3}$ down to $10^{-3}\, \cm^{-3}$.
We obtain $3\mu$G$\le B_{\rm eq}\le 30\mu$G.

\begin{figure}
\epsfxsize=\hsize \epsfbox{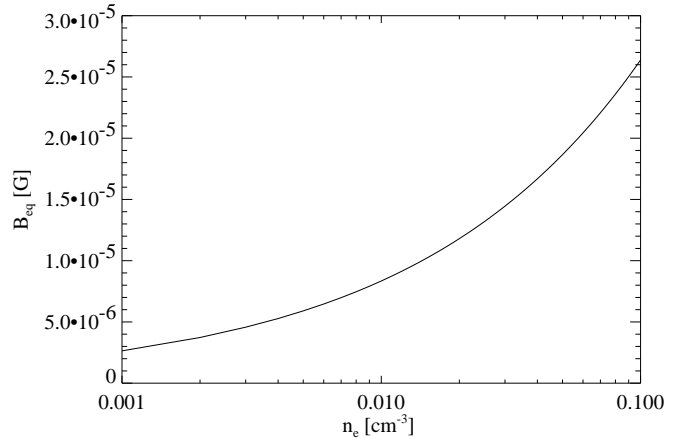}
\caption[]{Magnetic field strength $B_{eq}$ equivalent to the temperature
$2\cdot10^6 \K$ vs. the number density.
}
\end{figure}

Next, we calculate the magnetic heating rate
\begin{equation}
Q = \frac{B^2}{8\pi} \epsilon v_{rec} A
\end{equation}
where $A$ is the surface area of the interface of two flux tubes, $v_{rec} =
V_A/\ln(N_L)$ is the reconnection rate and $\epsilon$ is the efficiency of
magnetic energy conversion into heat. Let us suppose that $\epsilon = 1/2$.
If the flux tube diameter is $D$ of the order of 10 pc, then the surface area
at the interface of two flux tubes is of the order of $A = 100 \pc^2$.  We
assume here that the typical configuration of colliding flux tubes is similar
to that presented in Fig.~1, where the collision regions are marked with
circles.  We can estimate the heat released by a single reconnection event i.e.
by the complete reconnection of two flux tubes with the cross section area
$A\sim 100 \pc^2$ and the magnetic field $B\sim 10 \mu\G$.  Following our
estimation of $N_L\sim 5\cdot10^{12}$ one obtains $\ln (N_L) \sim 30$.  The
heating rate during the single reconnection event is

\begin{equation}
Q_1 = 1.2\cdot 10^{33} \left[\frac{B}{10\mu\G}\right]^3
                  \b \left[\frac{n_e}{0.01\cm^{-3}}\right]^{-\frac{1}{2}}
                  \b \left[\frac{A}{100 \pc^2}\right]\b \erg/\s
\end{equation}
The duration of the reconnection event is
\begin{eqnarray}
t_{rec} &=& \frac{D}{v_{rec}} \label{t_rec}  \\
        &=& 1.5 \left[\frac{D}{10 \pc}\right]
           \left[\frac{B}{10\mu\G}\right]^{-1}
     \left[\frac{n_e}{0.01 cm^{-3}}\right]^{\frac{1}{2}}  \Myr \nonumber
\end{eqnarray}
The estimated reconnection times can be confronted with observations.
Since the magnetic field in the arms of NGC 6946 is turbulent in arms and
uniform in the interarm regions (Beck and Hoernes, 1996) the time available for
the postulated uniformization process, sketched in Figs.~1 and 2, is as short
as the quarter of rotation period.  We note that the described process
eliminates the irregularities introduced by the Parker-shearing instability.
In the final state the major part of the nonuniform magnetic field is
replaced by the new radial magnetic field component.

The mentioned quarter of the rotation period is approximately the time
which elapses between the passages of  subsequent maximum and minimum of
density corresponding to the spiral density wave. This estimation can be done
more precisely, however we can assume a typical galactic rotation period
$t_{rot} \sim 200 \Myr$, which leads to a rough limitation
\begin{equation}
t_{rec} \leq 50 \Myr,
\end{equation}
if the magnetic arms in between the optical arms are to be formed.
Since the reconnection is supposed to
undergo in the upper layers of the galactic thin disc we can assume electron
densities down to $n_e \sim 10^{-2} \cm^{-3}$, so the reconnection times for
single flux tube can be of the order of a few Myr.  This means that we can
consider flux tubes much thicker than the assumed diameter 10 pc, or
alternatively domains containing many coherently moving flux tubes.  The above
limitation on the reconnection time seems to allow for the domain size
transversal to the mean magnetic field up to 100 pc.  This appears to be
consistent with the typically adopted space scale of the interstellar
turbulence.  It is important to notice that the reconnection time is
proportional  to square root of density,  which implies that reconnection goes
faster for higher heights over the galactic symmetry plane.
If the reconnection is not sufficiently efficient at a specific height, the
rising buoyant motion makes it easier and easier.
Summarising, we can say that magnetic reconnection is able to
remove irregularities of magnetic field up to the scale of 100 pc
in between the passages of subsequent arm and interarm region.

This conclusion supports the galactic dynamo model proposed in the Paper I,
which represents a  cyclic process related to the density
wave cycle.  In that model the magnitude of differential force (dependent on
the phase of density wave) controls the magnitude of the $\alpha$-effect.
Additionally, from Paper I we know that both the growth rate of the
Parker-shearing instability and the $\alpha$-effect are strongly dependent on
the cosmic ray pressure.  The cosmic rays are accelerated in the supernova
remnants in the rate dependent again on the phase of the density wave.  Due to
this fact the instability is the most vigorous in arms, making the magnetic
field nonuniform.  Then reconnection takes place and a huge amount of cosmic
rays decouples from the disc together with the upper family of magnetic field
lines.  The lower family of magnetic field lines which remains in the interarm
regions of the disc is after reconnection less populated by cosmic rays, more
stable and in consequence more uniform than the magnetic field in arms.

The total energy released during the 1 reconnection event is
\begin{eqnarray}
E_1 &=& Q_1 \ t_{rec} = \frac{B^2}{8\pi} \epsilon D^3 \\
    &=& 5.4 \cdot 10^{46} \left[\frac{B}{10\mu\G}\right]^2
                    \left[\frac{D}{10 \pc}\right]^3 \erg   \nonumber
\end{eqnarray}
The number of reconnection events can be estimated as follows. Let us suppose
that the galactic disc radius is 20 kpc and the disc full thickness is 500 pc.
If the flux tube volume filling factor is close to 1 as we assumed in Paper I
then the disc could contain $ \sim 2000 \cdot 50=10^5$ ringlike flux tubes
with the typical diameter 10 pc.  In fact we do not assume that the flux tubes
are rings going around the entire disc since there are pitch angles up to 30
degrees.  The above picture is convenient, however, for the current order of
magnitude estimations.  The average length of flux tubes is $L = 2\pi \cdot10
\kpc = 60 \kpc$.  Depending on the amount of cosmic rays the typical
wavelengths of the Parker instability is $\lambda=0.1-1 \kpc$.  The upper limit
is for the equipartition of the energies of cosmic ray gas and the magnetic
field, and the lower limit is for the significant excess of cosmic rays, what
has been discussed in the Papers I and II. Then the full range of the
azimuthal angle $\varphi$ corresponds on average to 60-600 spatial periods of
the waves associated to the Parker-shearing instability. The dynamo model
described in the Section~2 requires one reconnection event per spatial period
of undulations.  The fulfillment of this condition would ensure the maximal
efficiency of the generation of magnetic field, however some smaller numbers of
reconnection events would be more realistic.  Let us fix our attention on the
maximal value.  {}From our estimations it follows that the assumed disc
contains

\begin{equation}
N = 6\cdot (10^6 - 10^7)\cdot\left[\frac{D}{10 \pc}\right]^{-2}
\end{equation}
spatial periods of the instability and as many reconnection events should take
place within $T_{rot}/2$ assuming that the onset of the Parker-shearing
instability is triggered by the enhanced production of cosmic rays by
supernovae in spiral arms.  This assumption implies that the reconnecting loops
of magnetic field are formed once in between arms.  As it has been already
mentioned, we imagine that the cosmic rays are first injected by supernova
explosions, then they are raised by the Parker-shearing instability and finally
they are disconnected by the magnetic reconnection.  The next sequence of the
three above processes can be triggered only by the next spiral arm.  The
total heating rate of the galactic disc coming from reconnection is then

\begin{eqnarray}
\lefteqn{Q_{tot} = \frac{2 N E_1}{t_{rot}}} \label{Q_total}\\
&&   =(10^{38}-10^{39}) \left[\frac{B}{10\mu\G}\right]^2
                        \left[\frac{D}{10 \pc}\right]
                    \left[\frac{t_{rot}}{200 \Myr}\right]\erg/\s  \nonumber
\end{eqnarray}
Let us note that the above heating rate results from dissipation of
a fraction of the total (as well as turbulent) magnetic energy of the disc
\begin{equation}
f= \frac{\epsilon D^3}{\lambda D^2} E_{m\,tot}
=\frac{D}{\lambda} E_{m\,tot},
\end{equation}
which is 1 to 10 \% for the current choice of parameters.  This implies that
only a small fraction of the energy of the nonuniform component of magnetic
field is converted to the heat. However, the process described in the Figs.~1
and 2 efficiently removes the irregularities of magnetic field.
The remaining part of vanishing turbulent magnetic energy is first used to
built the new configuration of magnetic field and then partially expelled
from the disc together with the upper family of magnetic field lines. For this
reason the term relaxation is more proper for the described process than
the dissipation.  On the other hand, the dynamo models which require the full
dissipation of the small scale component would produce one to two orders of
magnitude
larger heating rates.  Since the results of Beck and Hoernes (1996) allow to
estimate the uniformization time, the amount of turbulent magnetic energy,
which decays between arm and interarm regions within that time, the rate of
such relaxation or dissipation processes can be compared with observations of
the diffuse X-ray emissivity of galaxies (see the next sections of the paper).
Such comparison can serve as a test of currents dynamo models.

Since the heating is associated with temperatures $3\cdot 10^6 \K$ we expect
the heat conversion to the X-ray luminosity via cooling processes.
The cooling process of a very hot plasma  with temperature more than $10^7$ K
is thermal bremstrahlung or free-free emission. Gas at lower temperatures cools
mainly by electron impact excitation of electronic levels of the neutral and
ionized particles. Dalgarno and McCray (1972) derived the interstellar cooling
function ${\cal L}$  which is at its maximum of the
order of $10^{-21}\, \erg \s^{-1} \cm^3$ (at about $10^4$K) and about
$2\cdot 10^{-23}\, \erg \s^{-1} \cm^3$ at temperatures above $10^6$K.
We can compare the magnetic heating time with the cooling time.
The heating time is
\begin{equation}
t_{heat} \equiv \frac{n k_B T}{q},
\end{equation}
where
\begin{equation}
q \equiv Q_1/D^3
\end{equation}
is the heating rate per unit volume in the reconnection region of the volume
$D^3$. After substitutions we obtain
\begin{equation}
t_{heat}=1\left[\frac{n_e}{0.01 \cm^{-3}}\right]
       \b   \left[\frac{T}{10^6 \K}\right]
       \b   \left[\frac{D}{10 \pc}\right]
       \b   \left[\frac{B}{10 \mu\G}\right]^{-3} \b\Myr. \label{t_heat}
\end{equation}

The cooling rate (per unit volume ) and the cooling time are respectively
\begin{equation}
\Lambda\equiv n^2\cdot {\cal L},
\end{equation}
and
\begin{eqnarray}
\lefteqn{t_{cool}\equiv \frac{n k_B T}{\Lambda}}  \label{t_cool}  \\
&&        =  50 \left[\frac{T}{10^6 \K}\right]
          \left[\frac{{\cal L}}{10^{-23} \frac{\erg \cm^3}{\s}}\right]^{-1}
          \left[\frac{n}{0.01 \cm^{-3}}\right] \Myr \nonumber
\end{eqnarray}
The three relevant time scales $t_{rec}$, $t_{heat}$ and $t_{cool}$ resulting
from formulae (\ref{t_rec}), (\ref{t_heat}) and (\ref{t_cool}) respectively are
shown in Fig.~4 for different values of the magnetic field strengths $B$ and
the flux tube diameters (or domain sizes) $D$.

\begin{figure*}
\epsfxsize=\hsize \epsfbox{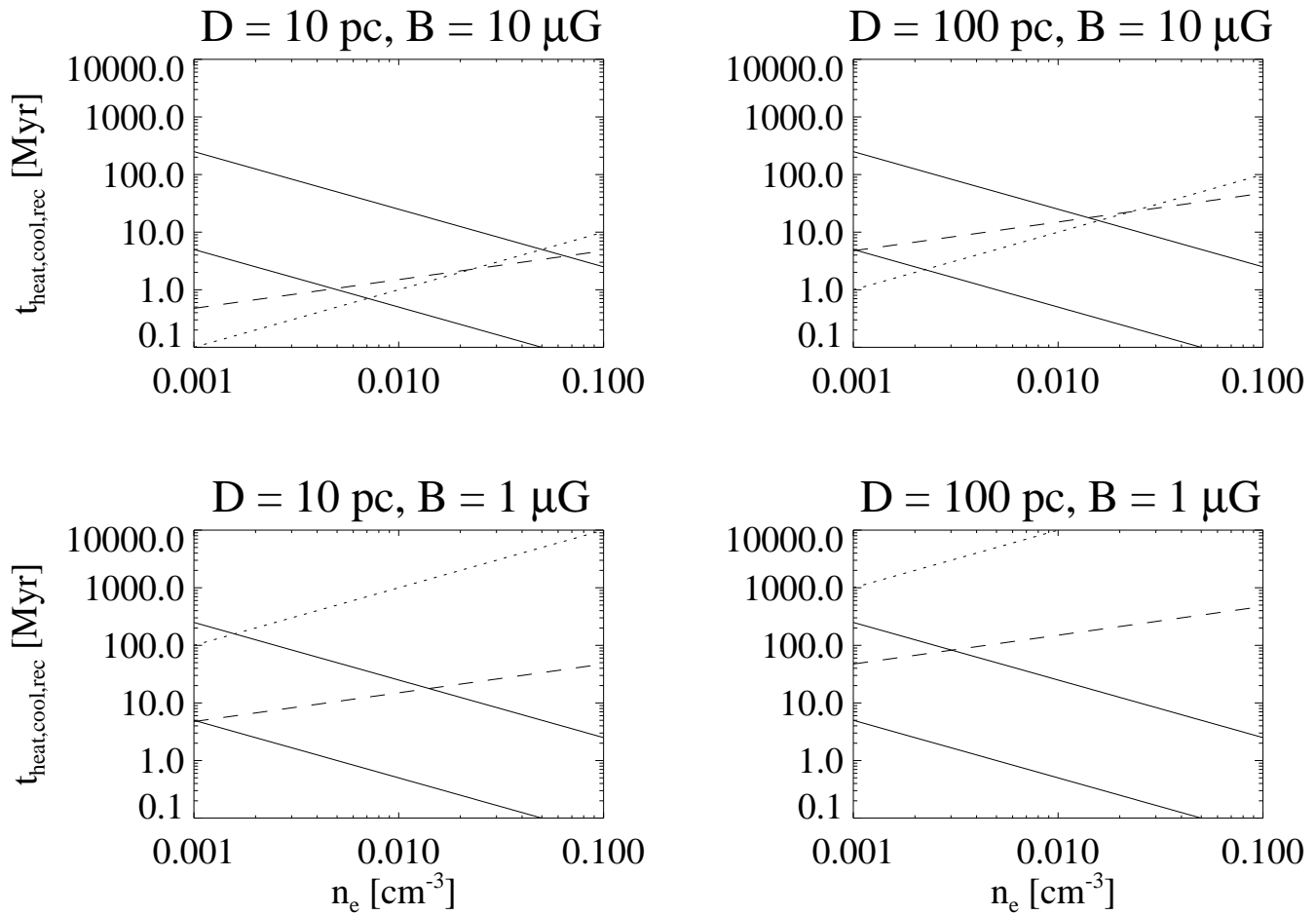}
\caption[]{
The three relevant time scales $t_{rec}$, $t_{heat}$ and $t_{cool}$ resulting
are shown  for various values of the magnetic field strengths $B$ and
the flux tube diameter (or domane size) $D$. For all panels $T=10^6 \K$, two
characteristic values of the cooling function are represented by the upper full
line (${\cal L} = 2\cdot10^{-23} \erg \cm^{-3} \s^{-1}$) and the lower full
line (${\cal L} = 10^{-21} \erg \cm^{-3} \s^{-1}$). The heating
time is drown with the dotted line and the reconnection time with the dashed
line.}
\end{figure*}

We note in Fig.~4 that the strength of magnetic field determines the relation
between the reconnection and heating timescales. For $B\sim 10\mu\G$ both the
timescales are comparable, and the heating timescale is typically shorter than
the reconnection timescales for $n_e \leq 0.01 \cm^{-3}$ (higher levels
of the galactic disc/halo system). This implies that at higher levels of the
galactic disc the heating process leads to high temperatures in the
single reconnection region during the reconnection time. The cooling
timescales related to two characteristic values of cooling functions
$10^{-21}$ and $2\cdot10^{-23}\erg\cm^3\s^{-1}$ are represented by two full
lines in all panels.  The upper line is for slower cooling typical for higher
temperatures (above $10^6$ K) and the lower line is for faster cooling typical
for lower temperatures (below $10^6$ K).  It is remarkable that strong magnetic
fields, densities below $10^{-2} \cm^{-3}$ and moderate flux tube diameters (or
domain sizes) allow the interstellar gas to be heated from cool initial states
because the heating time is shorter than the shorter of the cooling times.  In the
case of larger structures the same is possible if $n\leq10^{-3} \cm^{-3}$ or
the heating process starts from initial temperatures $T\sim 10^6 \K$.  Such
initial temperatures can be attained in many different ways, e.g.  in former
reconnection events, due to locally very concentrated magnetic structures,
streaming of cosmic rays along magnetic field or a supernova contributions.

If magnetic field is weaker, e.g. $B \sim 1\mu\G$, the heating timescale is
much longer than the reconnection timescale due to different powers of magnetic
field in formulae (\ref{t_rec}) and (\ref{t_heat}): $t_{rec} \sim B^{-1}$ and
$t_{heat} \sim B^{-3}$ respectively. Moreover, the heating timescale starts to
exceed the slower cooling timescale at $n_e \sim 10^{-3} \cm^{-3}$, so
reconnection is able to heat gas to X-ray emitting temperatures for densities
smaller than the mentioned above (in upper levels of galactic halos) in
multiple reconnection events, or there is no X-ray emission resulting from
reconnection in some cases.  The last case however, does not mean a lack of
reconnection itself, but only a lack of the associated X-ray emission.  One
should mention in this context that the other heating processes can still work
in such condition and overwhelm the heating by reconnection.

For weaker magnetic fields $\sim 1\mu\G$ the reconnection timescale, although
shorter than the heating timescale is growing inversely to the magnetic field
strength. This suggests that in the limit of very  weak magnetic fields the
reconnection time becomes significantly longer than the galactic dynamical
timescale, what would imply that our dynamo model fails to work.
One should remember however, that in the range of high plasma betas, magnetic
field tends to be more and more intermittent i.e. concentrated in isolated flux
tubes. The magnetic field evolution in such a flux tube is then completely
determined by the flux tube dynamics.
The field concentration should ensure that  magnetic
field strength within magnetic flux tubes is independent on the total magnetic
energy of galactic discs. This in turn would imply an efficient dissipation by
reconnection even in the weak (volume average) magnetic field limit.

In the expression (\ref{Q_total}) we estimated the total heat output from
reconnection in the volume of galactic disc, which typically should be of the
order of $10^{38} - 10^{39} \erg/\s$ for galaxies similar to ours or NGC 6946.
The heat released is converted to the X-ray luminosity via the cooling
processes as described by Dalgarno and McCray (1972). The X-ray luminosity
can be computed by multiplying the cooling function by the X-ray emiting
volume
\begin{equation}
P_X = {\cal L} n^2 V_{disc} \eta,
\end{equation}
where $V_{disc}$ is the total volume of galactic disc and $\eta$ is the volume
filling factor of the X-ray emiting gas. One can estimate
\begin{equation}
P_X =  1.5 \cdot 10^{40} \left[\frac{n}{0.01\cm^{-3}}\right]^2
                       \b \left[\frac{H}{0.5 \kpc}\right]
                        \b\left[\frac{R}{20 \kpc}\right]^2
                        \b\eta \frac{\erg}{\s},
\end{equation}
where $H$ is the disc (full) thickness and $R$ is the disc radius. The volume
filling factor of X-ray emitting gas is a poorly determined parameter.
On the other hand the X-ray luminosity resulting
from reconnection should not exceed the amount of heat released by
reconnection. This gives an upper limit of $\eta$ for the reconnection heated
X-ray emiting gas
\begin{equation}
\eta_{rec} \leq 0.1
\end{equation}
Finally we would suggest that if $t_{heat}$ is shorter than $t_{cool}$, a
majority of released heat is converted to the X-ray luminosity, which according
to our estimations should be $10^{38} - 10^{39} \erg/\s$ for galaxies like ours
or NGC 6946.

\section{Comments on the X-ray emission of selected galaxies}

Schlegel (1994) analyzes the X-ray ROSAT observations of NGC 6946.  The X-ray
image of that galaxy is composed of a number of point sources and a diffuse
emission.  Schlegel concludes that the diffuse emission is not the sum of
unresolved weak point sources like supernova remnants and X-ray binaries.  We
suggest that the unresolved X-ray emission can be related to magnetic heating
process described above.

In M51 (Ehle, Pietsch and Beck, 1995) an X-ray emission comes from outside the
optical limits and especially from the intergalactic region between M51 and its
companion.  This emission is hard to explain by the presence of binary system.
As in the case of NGC 6946 a significant part of X-ray emission is not resolved
into individual point sources.  The authors argue that the diffuse emission can
not be explained by a large number of unresolved point sources.  The scale
lengths of the thermal radio continuum emission and of the diffuse X-ray
emission are similar.  Although it is suggested that a hot gas from starburst
regions (transported via galactic fountains or winds) can be responsible for
the diffuse emission, we point out that the Parker-shearing instability
together with the magnetic reconnection can lead to the same effect.  It is
worth noting (see Fig.  1.  of Ehle, Pietsch and Beck (1995)) that the X-ray
emission of M51 is locally enhanced in the wide interarm region internal to the
northern spiral arm.  This effect may be explained by the magnetic
reconnection, which according to our expectations, should extend toward
interarm regions and take part in the regularization of the magnetic structure
as we suggest in Papers I and II.  In addition, energy densities of magnetic
field and hot gas are comparable in galactic halo of M51, which may follow from
direct conversion from magnetic to thermal energy due to the magnetic
reconnection.  There is a similar effect in NGC 1566 (Ehle et al.  1996),
where asymmetrically extended X-ray emission flares out from the nucleus
in the direction toward the two interarm regions of enhanced polarized
radio emission.

\section{ Discussion and conclusions}

The main result of the present paper is the new dynamo model following from
our previous results of Papers I and II. Our model is a kind of fast
galactic dynamo proposed earlier by Parker (1992). The main difference
between the Parker's model and ours results from the remark that due to the
cosmic rays, differential force and the density waves, the magnetic
tension has only a little influence on the flux tube dynamics.
The cyclonic deformation of flux tubes due to the Parker-shearing
instability is strong in both the vertical and radial directions. Moreover,
the radial deformations  exceed the cosmic ray thickening of
the upper parts of flux tubes. This implies that the magnetic reconnection
operates preferentially at places where two different flux tubes are
pressed together in the radial direction by the Coriolis force.
With the aid of magnetic reconnection this process
(see Figs.~1 and 2)  generates two new families of flux tubes from the
buckled initial family. Both the families
contain new components of the opposite sign, large scale  radial magnetic field.
The upper family, twisted with respect to the original
magnetic field in direction corresponding to
the Coriolis force, forms at lower parts of galactic halo, is inflated by
cosmic rays and is lost from the disc. The lower family is twisted in the
direction opposite to the Coriolis force and forms in the disc.
In this way the sign of the $\alpha$-effect in the disc is negative
and the dynamo number $D$ is positive. This circumstance favors the
solutions of the dynamo equation which are propagating and are able to couple
with density waves providing an additional growth rate to the dynamo waves.
The negative $\alpha$-effect and the differential rotation considered separately
lead to the same sign of the magnetic pitch angle in the disc as the sign of
the pitch angle of spiral arms. In the halo, on the other hand the rotation of
magnetic field lines corresponding to the $\alpha$-effect is opposite to the
effect of differential rotation. Since the Parker-shearing instability may be
in some regions of the galactic discs much more vigorous
than in other places, we expect large variations and reversals of the magnetic
pitch angle in the adjacent regions of galactic halos.  It is a specific feature
of our model that the $\alpha$-effect in disc and halo are closely related and
have opposite signs.  We would like to point out that the negative
$\alpha$-effect in galactic discs, resulting from the model presented in this
paper  is not inconsistent with the former Papers I and II. In those papers we
calculated the helicity applying the standard formulae. In the present paper we
have introduced the new scenario of flux tube interactions, which serves as a
physical interpretation of the former formal results. According to this
scenario the $\alpha$-effect resulting from both the Parker-shearing
instability and the subsequent relaxation process is negative in discs
and positive in halos.

In the frame of our model we estimated the rate of generation of the radial
component as well as the loss rate of magnetic field from the disc.  The
transport coefficients derived on the base of simple geometrical considerations
led us to the expressions which are equivalent to the formally derived helicity
$\alpha_d$ and diffusivity $\eta_d$ in Paper I.

Since our model involves magnetic reconnection, which is responsible for the
relaxation of magnetic field structure, the dynamo process we describe
implies  the magnetic heating of interstellar gas. The estimated X-ray luminosity
of spiral galaxies of the order of $10^{38} - 10^{39}$ erg/s is comparable,
to the observed X-ray luminosities of selected nearby galaxies.
We would like to point out that in our model only 1 to 10 \% of the
magnetic turbulent energy is directly dissipated (i.e.  converted to the heat)
in the described dynamo process by means of the magnetic reconnection.  The
rest of turbulent energy is replaced, in the described relaxation process, by
the large scale component of magnetic field, which in turn is partially
expelled to the halo and then to intergalactic space.

Our estimation of X-ray luminosity is not precise, arising
as a byproduct of the proposed dynamo model.  This X-ray luminosity
does not take into account an amount of magnetic reconnection which does not
directly take part in the dynamo process, but only smoothes magnetic field
irregularities on single, strongly inflated flux tubes in halo.  We expect that
in some extreme cases of high cosmic ray production rate, the mentioned
contribution can be even stronger by more than two orders of magnitude, because
of dominating vertical structures (lobes) of magnetic field with opposite
polarity extending to the heights of a few kpc.  The associated total surface
area of current sheets should be extremely large and conditions (mainly low
density) very favorable for an efficient reconnection leading to the X-ray
emission up to $10^{42} \erg/\s$.

\begin{acknowledgements}
Prof.  Eugene Parker is greatly acknowledged for the careful reading,
improvements and discussion on the manuscript.  The basic ideas of the paper
were formulated during the workshop on "Galactic and Cosmological magnetic
fields" at the Aspen Center for Physics in summer 1996.  This work was
supported by the grant from Polish Committee for Scientific Research (KBN),
grant no.  2P03D 016 13 and the Deutsche Forschungsgemeinschaft through the
grant ME 745/18-1

\end{acknowledgements}

\end{document}